# Templates for magnetic symmetry and altermagnetism in hexagonal MnTe

S. W. Lovesey[1,2,3], D. D. Khalyavin[1] and G. van der Laan[2]

[1]*ISIS Facility, STFC, Didcot, Oxfordshire OX11 0QX, United Kingdom*
[2]*Diamond Light Source, Harwell Science and Innovation Campus, Didcot, Oxfordshire OX11 0DE, United Kingdom*
[3]*Department of Physics, Oxford University, Oxford OX1 3PU, UK*

**Abstract** The symmetry of long-range magnetic order in manganese telluride ($\alpha$-MnTe) is unknown. Likewise, its standing as an altermagnet. To improve the situation, we present symmetry informed Bragg diffraction patterns based on a primary magnetic order parameter for antiferromagnetic alignment between Mn dipoles. It does not break translation symmetry in a centrosymmetric structure, in keeping with an accepted definition of altermagnetism. Magnetic symmetry forbids a linear magnetoelectric (ME) effect and allows a piezomagnetic (PM) effect. The proposal is to check diffraction patterns measured in the future against four templates that differ with respect to the orientation of Mn dipoles in the basal plane and canting out of the plane. Our templates serve x-ray diffraction that benefits from signal enhancement using a Mn atomic resonance, and neutron scattering. Even rank multipoles in magnetic neutron diffraction reflect a core requirement of altermagnetism, because they are zero for strong spin-orbit coupling. Symmetry in the templates demands that nuclear and magnetic contributions possess the same phase, which enables standard neutron polarization analysis on Bragg spots with overlapping contributions. However, three of the four templates generate Bragg spots that do not appear in the lattice (nuclear) diffraction pattern, i.e., Bragg spots that are basis-forbidden and purely magnetic in origin. On the other hand, identical symmetry demands a 90º phase shift between magnetic (time-odd) and charge-like (time-even, Templeton-Templeton) contributions to x-ray scattering amplitudes. Consequently, circular polarization in the primary beam of x-rays is rotated. The difference in the intensities of a Bragg spot measured with right- and left-handed circular primary polarization defines a chiral signature, much like magnetic circular dichroism (XMCD). Bragg spot intensities and the chiral signal may change with rotation of the illuminated crystal about the reflection vector, depending on the template and the placement of the reflection vector in the reciprocal lattice. Further tests include predictions in three out four templates of zero intensity in a specified channel of x-ray polarization. Diffraction properties of a template are radically different from those of a parity-time (PT)-symmetric antiferromagnet, for its symmetry allows a linear ME effect and prohibits both a PM effect and a chiral signature.

## I. Introduction

Early studies of hexagonal manganese telluride ($\alpha$-MnTe) included the use of fixed solutions ZnTe/MnTe and thin films [1, 2]. It is a correlated magnetic semiconductor with a moderate indirect band-gap $E_g \approx (1.27–1.46)$ eV, and long-range magnetic order below a Néel temperature 307 - 310 K [3, 4]. A piezomagnetic (PM) effect, and neutron Bragg diffraction and inelastic scattering (spin waves), and magnetic susceptibility measurements have been reported for powder samples and single crystals [3, 4, 5, 6]. By and large, however, recent measurements are performed on MnTe films [7, 8, 9]. The promotion of MnTe as a candidate for bulk altermagnetism has led to a surge of publications about its magnetic and transport properties [10-15]. An accepted definition of an altermagnet includes the absence of spin-orbit coupling and presence of non-relativistic collinear antiferromagnetism. Moreover, in this definition the two sublattices are related by symmetries other than translation or inversion symmetry [12]. In the absence of a consensus for the magnetic symmetry of MnTe crystals discussions of its altermagnetism inevitably remain speculative. For example, a PM effect is interpreted with a crystal class mmm [6], and discussions of magnetic circular dichroism (XMCD) measurements and simulations proceed with m'm'm [9]. In light of the current uncertainty for the magnetic symmetry of MnTe, we study templates of magnetic symmetry that fit the aforementioned definition of altermagnetism. Our four templates are derived from the parent NiAs-type structure of MnTe, and their essential properties are gathered in Table I. Corresponding x-ray and magnetic neutron scattering amplitudes (Sections V and VI) to be tested in future experiments are inferred from crystal and magnetic symmetry.

The x-ray scattering we discuss uses a Mn atomic resonance to enhance the intensity of Bragg spots. This attribute enables x-ray diffraction to measure more weak magnetic reflections that are forbidden by the symmetry of the parent lattice. A non-magnetic analogue are Bragg spots in an x-ray diffraction pattern created by departures from spherical symmetry of electronic charge: spherical scatterers on a regular lattice define a parent space group. Polarization analysis is a second attribute of resonant x-ray diffraction that we explore. All our templates rotate the helicity (handedness) of primary x-rays. A difference in intensities of a Bragg spot measured with left- and right-handed primary x-rays defines a chiral signature as a quantitative measure of the rotation of the plane of polarization. Our x-ray diffraction amplitudes and chiral signatures allow for rotation of the illuminated crystal through an azimuthal angle about the reflection vector. Chiral signatures hallmark magnetic symmetry in the templates. In an Appendix, we show they are forbidden in a parity-time (PT)-symmetric antiferromagnet, for example, that supports a linear magnetoelectric (ME) effect and forbids a PM effect. By way of an example of the information provided by an azimuthal angle scan we anticipate the result for chiral signatures of orthorhombic templates (1) and (2). Signatures are two-fold symmetric functions of the azimuthal angle when the reflection vector is parallel to the crystal c axis and thus the dyad axis of rotation symmetry. Magnetic neutron diffraction is the most widely used technique by which to establish the magnetic structure of a material. By and large, diffraction patterns are interpreted in terms of magnetic dipoles alone. Beyond are multipoles of higher rank that exist for particular electronic configurations. Even rank multipoles in magnetic neutron diffraction exist when the atomic configuration includes two or more J-states. Such is the case for zero, or very weak, spin-orbit coupling in the definition of altermagnetism. To be specific, the neutron quadrupole measures the entanglement of a spin anapole and orbital degrees of freedom, and the hexadecapole has a similar equivalence.

**II. Structure properties**

The parent structure of bulk MnTe is P6$_3$/mmc (No. 194, crystal class 6/mmm [16]), with Mn$^{2+}$ (3d$^5$) and Te$^{2-}$ ions in sites 2a (0, 0, 0) and 4f (1/3, 2/3, 1/4), respectively, with cell dimensions $a \approx 4.190$ Å, $c \approx 6.751$ Å [4]. Manganese Mn$^{2+}$ is an s-state ion with spin S = 5/2 and orbital angular momentum L = 0. Unit cell vectors are **a**$_h$ = ($a$, 0, 0) parallel to a dyad axis of rotation symmetry normal to the triad c axis, and **b**$_h$ = (1/2) (− $a$, $a$√3, 0) and **c**$_h$ = (0, 0, c), that enclose a unit cell volume v$_o$ = (1/2) $a^2c$√3.

### III. Magnetic symmetries

An antiferromagnetic alignment between axial dipoles of the two Mn ions is the primary order parameter common to all four templates discussed here. Symmetries therein differ with respect to directions of the dipoles canting out of the basal plane, cf. Table I. Choice of the cells follows standard crystallographic requirements. They respect the translation symmetry of the crystal (including both structural and magnetic degrees of freedom), have minimal unit cell volume, and represent the standard setting for the magnetic space groups [16]. Magnetic symmetries are centrosymmetric, possess a magnetic order that does not break translation symmetry [a propagation vector **k** = (0, 0, 0)], and anti-inversion is absent in the magnetic crystal class. Absence of anti-inversion precludes a linear ME effect, of course, and Landau free-energies include a contribution HEE, where H and E denote magnetic and electric fields. All templates allow a PM effect [6]. Ferromagnetism parallel to the c axis is a secondary order parameter in three templates (it is expected to be vanishingly small).

Looking at Table I, our four templates in brief are: (1) Cm'c'm (No. 63.462 (BNS [16]), magnetic crystal class m'm'm); (2) Cmcm (No. 63.457, mmm); (3) P2$_1$/m (No. 11.50, 2/m); (4) C2'/m' (No. 12.62, 2'/m'). The Néel vector is confined to the basal plane in (3), while the demands of additional symmetries in (1) and (2) place it parallel to (1, −1, 0)$_h$, i.e., off-set by 30° from **a**$_h$, and **a**$_h$, respectively. Symmetries in template (4) permit dipoles to cant out of the basal plane, with axial moments along (1, −1, 0)$_h$, as in (1), and the c axis. Manganese ions occupy sites that are centres of inversion symmetry.

A net magnetic field allows an XMCD signal for x-rays travelling along the c axis, while XMCD = 0 for x-rays propagating in the basal plane [17, 18, 19]. Experiments by Hariki *et al*. do not reveal an XMCD signal at Mn L$_{2,3}$ edges in a MnTe film contrary to expectations derived from their simulations of an electronic structure compatible with templates (1) and (4) [9]. Regarding resonant x-ray and neutron diffraction amplitudes for magnetic symmetries, which are the principal results from our study, the former are complex and the latter can be chosen purely real without loss of generality [20-24, 25-27]. In consequence, an x-ray chiral signature is permitted and the intensity of a Bragg spot is the sum of squares, one component comprised of magnetic multipoles and one component comprised of charge-like (Templeton-Templeton (T & T) [20]) multipoles. Chiral signatures are a product of magnetic and charge-like multipoles, cf. Table II. The common phase of nuclear and magnetic contributions to neutron diffraction amplitudes permits an interference of the two components routinely exploited in neutron polarization analysis [28, 29].

For an atomic description of electronic degrees of freedom, Mn ions are assigned spherical multipoles $\langle O^K_Q \rangle$ as in Table II. They encapsulate properties of the electronic and magnetic ground state of MnTe. Cartesian and spherical components $Q = 0, \pm 1$ of a vector $\mathbf{n} = (\xi, \eta, \zeta)$, for example, are related by $\xi = (n_{-1} - n_{+1})/\sqrt{2}$, $\eta = i(n_{-1} + n_{+1})/\sqrt{2}$, $\zeta = n_0$. A complex conjugate of a multipole is defined as $\langle O^K_Q \rangle^* = (-1)^Q \langle O^K_{-Q} \rangle$, meaning the diagonal multipole $\langle O^K_0 \rangle$ is purely real. The phase convention for real and imaginary parts labelled by single and double primes is $\langle O^K_Q \rangle = [\langle O^K_Q \rangle' + i\langle O^K_Q \rangle'']$. Whereupon, $\langle O^1_\xi \rangle = -\sqrt{2} \langle O^1_{+1} \rangle'$ and $\langle O^1_\eta \rangle = -\sqrt{2} \langle O^1_{+1} \rangle''$.

Multipoles for each template are specified in local orthogonal coordinates $(\xi, \eta, \zeta)$ and they are,

$$(1) \quad \xi_1 = (1, 0, 0)_h; \eta_1 = (1, 2, 0)_h; \zeta_1 = (0, 0, 1)_h; \tag{1}$$

$$(2) \quad \xi_2 = (1, 1, 0)_h; \eta_2 = (-1, 1, 0)_h; \zeta_2 = (0, 0, 1)_h. \tag{2}$$

Here, $(1, 2, 0)_h \equiv (1, -1, 0)_h \propto (0, 1, 0)$ and $(1, 1, 0)_h \equiv (1, 0, 0)_h \propto (1/2)(1, \sqrt{3}, 0)$ where the equivalences exploit the triad axis of rotation symmetry along the c axis. With unique axis $\eta_3$,

$$(3) \quad \xi_3 = -(2, 1, 0)_h \propto -(1/2)(\sqrt{3}, 1, 0); \eta_3 = (0, 0, -1)_h; \zeta_3 = (0, -1, 0)_h. \tag{3}$$

In line with standard practice, $\xi_3$ is proportional to the reciprocal lattice vector $(\eta_3 \times \zeta_3)$. Likewise, in (4),

$$(4) \quad \xi_4 = -(1, 2, 0)_h; \eta_4 = (1, 0, 0)_h; \zeta_4 = (0, 0, 1)_h. \tag{4}$$

### IV. Electronic structure factors

An electronic structure factor,

$$\Psi^K_Q = [\exp(i\boldsymbol{\kappa} \cdot \mathbf{d}) \langle O^K_Q \rangle_\mathbf{d}], \tag{5}$$

determines neutron and x-ray diffraction patterns for a particular template. The implied sum in Eq. (5) is over positions $\mathbf{d}$ of Mn ions in a unit cell. The reflection vector $\boldsymbol{\kappa}$ is defined by integer Miller indices $(h, k, l)$ for all magnetic structures. For templates (1) and (2) one finds [16],

$$\Psi^K_Q(1, 2) = \langle O^K_Q \rangle [1 + (-1)^{h+k}][1 + (-1)^{l+Q}]. \tag{6}$$

Site symmetries impose conditions $\langle O^K_Q \rangle = [\sigma_\theta (-1)^K] \langle O^K_{-Q} \rangle$ and $\langle O^K_Q \rangle = (-1)^K \langle O^K_{-Q} \rangle$ for (1) and (2), respectively. The time signature $\sigma_\theta = +1$ for time even (charge-like or nuclear) and $\sigma_\theta$

= −1 for time odd (magnetic) multipoles, cf. Table II. In template (3) there are no additional constraints on $\langle O^K_Q \rangle$ beyond spatial inversion and [16],

$$\Psi^K_Q(3) = [\langle O^K_Q \rangle + (-1)^k (-1)^{K+Q} \langle O^K_{-Q} \rangle ]. \tag{7}$$

Two independent sites accommodate Mn ions in template (4), namely, 2a and 2c in C2'/m'. Both sites have symmetry 2'/m' that subjects multipoles to the constraint $\langle O^K_Q \rangle = [\sigma_\theta (-1)^{K+Q}] \langle O^K_{-Q} \rangle$. In consequence, magnetic dipoles are aligned with − (1, 2, 0)$_h$ and (0, 0, 1)$_h$.

Bulk properties of a material are determined by $\Psi^K_Q$ evaluated for $h = k = l = 0$. A selection rule for templates (1) and (2) is even Q. Specifically, a bulk property composed of dipoles is created by $\Psi^1_\zeta$, which is zero for template (2) on account of the constraint placed by site symmetry, while $\Psi^1_\zeta$ is allowed to be non-zero for magnetic dipoles in (1). The $\zeta$ axis is parallel to the c axis in both cases. The bulk value of $\Psi^1(3)$ can be non-zero on account of $\langle O^1_\eta \rangle$, and the unique monoclinic axis $\eta_3$ is parallel to the c axis. On the other hand, bulk value of $\Psi^1(4)$ are parallel to $\xi_4$ and $\zeta_4$. An XMCD signal is proportional to the projection of the net magnetic field $\Psi^1$ on the photon wave vector [17, 18, 19]. For the symmetries we consider, XMCD = 0 for the photon wavevector parallel to the basal plane and template (2).

### V. X-ray Bragg diffraction enhanced by E1-E1 or E2-E2 events

The triangle rule for vectors permits electronic multipoles of rank K = 0, 1 and 2 in an electric dipole-electric dipole (E1-E1) absorption event [18, 20]. Looking at Table II, dipoles (K = 1) are magnetic (time-odd), and multipoles with rank K = 0 and K = 2 are charge-like (time-even). Manganese absorption edges used in an E1-E1 scattering event have energies E ≈ 6.537 keV for the K edge (1s → 4p), and $L_2$ ≈ 0.649 keV and $L_3$ ≈ 0.638 keV (2p → 3d). The cell dimensions of MnTe are too small for diffraction enhanced by Mn L edges. There have been many rewarding experiments on 3d-transition metal compounds exploiting a K edge absorption event, and early examples include $V_2O_3$ and $TbMnO_3$ [21, 22]. Specifically, the Bragg angle in Fig. 3 is determined by ($\lambda$/2c), where the photon wavelength $\lambda$ ≈ (12.4/E) Å and the x-ray energy E is in units of keV. Whence, ($\lambda$/2c) ≈ 0.141 for MnTe and the Mn K edge. A resonance in an E1-E1 event is attributed to magnetic dipoles if it disappears above the Néel temperature. Furthermore, such a resonance contributes to the rotated channel of polarization and the chiral signature. The sensitivity to magnetic order at the 1s → 4p dipole transition-energy is due to the 4p − 3d intra-atomic Coulomb interaction and to the mixing of the 4p with the 3d states of neighbouring magnetic ions. At the E2 threshold, exhibiting five multipoles (K = 0 - 4), its origin is in the spin polarization of the 3d states. An E2-E2 amplitude is included in our discussion of template (4).

Diffraction amplitudes are labelled by states of primary and secondary polarization depicted in Fig. 3, e.g., ($\pi'\sigma$) is the energy-integrated diffraction amplitude for primary $\sigma$ polarization and secondary $\pi$ polarization. They are functions of the angle $\psi$ that measures rotation of the crystal about the reflection vector [18, 23, 24]. The start of an azimuthal angle scan $\psi = 0$ is defined by the placement of a lattice vector relative to the plane of scattering.

Consider lattice forbidden, or weak, Bragg reflections for which $\Psi^K_0 = 0$ with even K. Reflections are taken to be (0, 0, $l$) with odd $l$ for templates (1) and (2), and (0, $k$, 0) with odd $k$ for (3). For orthorhombic symmetries and a reflection vector (0, 0, $l$) the origin $\psi = 0$ is defined by $(-\zeta, -\eta, -\xi)$ in terms of coordinates (x, y, z) in Fig. 3. In consequence, vectors $\eta_1 = (1, 2, 0)_h \propto (0, 1, 0)$ and $\eta_2 = (-1, 1, 0)_h \propto (1/2) (-\sqrt{3}, 1, 0)$ are in the plane of scattering at the start of an azimuthal-angle scan for symmetries (1) and (2), respectively. Local coordinates $\eta_1$ and $\eta_2$ subtend an angle of 60º that appears as an off-set to relative azimuthal-angle scans. Corresponding x-ray dipoles are $\langle T^1_\eta \rangle_1$ and $\langle T^1_\xi \rangle_2$ and quadrupoles $\langle T^2_{+1} \rangle_{1,2}$ contributing to T & T scattering are related by changes to local coordinates [20]. Dipoles $\langle T^1_\xi \rangle_3$ and $\langle T^1_\zeta \rangle_3$ in directions $\xi_3 \equiv - (2, 1, 0)_h$ and $\zeta_3 \equiv - (0, 1, 0)_h$ subtend angles 150º and 60º with $a_h$.

Diffraction amplitudes for templates (1), (2) and (3) have a number of features in common. Unrotated amplitudes in the $\sigma$ channel are zero. With $(\sigma'\sigma) = 0$ a chiral signature,

$$\Upsilon = [(\pi'\pi)^*(\pi'\sigma)]''. \qquad (8)$$

In practice, $\Upsilon$ is the measured difference in intensities of a Bragg spot observed with oppositely handed primary x-rays. So, $\Upsilon$ and XMCD are alike with regard to polarization requirements. Unrotated amplitudes in the $\pi$ channel are magnetic and purely imaginary, while rotated amplitudes are a sum of magnetic (purely imaginary) and T & T (purely real) contributions. It follows that $\Upsilon$ is a measure of the interference between magnetic and charge-like degrees of freedom. Canting away from the basal plane allowed in template (4) yields distinctly different diffraction amplitudes, not least an absence of lattice forbidden reflections investigated for the remaining three templates.

*Template* (1) $\quad (\pi'\pi)_1 = i\sqrt{2} \sin(2\theta) \sin(\psi) \langle T^1_\eta \rangle_1, \qquad (9)$

$\quad (\pi'\sigma)_1 = \cos(\theta) \cos(\psi) [i\sqrt{2} \ \langle T^1_\eta \rangle_1 + 2 \langle T^2_{+1} \rangle_1''],$

$\quad \Upsilon(1) = -\sqrt{2} \sin(2\theta) \cos(\theta) \sin(2\psi) \langle T^1_\eta \rangle_1 \langle T^2_{+1} \rangle_1''.$

Reflection (0, 0, $l$) with odd $l$ and $\xi_1 = (1, 0, 0)_h$ normal to the plane of scattering at the start $\psi = 0$.

*Template* (2) $\quad (\pi'\pi)_2 = - i\sqrt{2} \sin(2\theta) \cos(\psi) \langle T^1_\xi \rangle_2, \qquad (10)$

$\quad (\pi'\sigma)_2 = \cos(\theta) \sin(\psi) [i\sqrt{2} \ \langle T^1_\xi \rangle_2 + 2 \langle T^2_{+1} \rangle_2''],$

$\quad \Upsilon(2) = \sqrt{2} \sin(2\theta) \cos(\theta) \sin(2\psi) \langle T^1_\xi \rangle_2 \langle T^2_{+1} \rangle_2''.$

Reflection $(0, 0, l)$ with odd $l$ and $\xi_2 = (1, 1, 0)_h$ normal to the plane of scattering at $\psi = 0$. The two-fold symmetry of $\Upsilon(1)$ and $\Upsilon(2)$ with respect to the azimuthal angle $\psi$ is due to dyad axes of symmetry in orthorhombic structures.

*Template* (3) $\quad (\pi'\pi)_3 = i\sqrt{2} \sin(2\theta) [\cos(\psi) \langle T^1\zeta\rangle_3 - \sin(\psi) \langle T^1\xi\rangle_3],$ $\qquad (11)$

$(\pi'\sigma)_3 = -i\sqrt{2} \cos(\theta) [\sin(\psi) \langle T^1\zeta\rangle_3 + \cos(\psi) \langle T^1\xi\rangle_3]$

$\qquad\qquad + 2 \cos(\theta) [\cos(\psi) \langle T^2_{+1}\rangle_3'' + \sin(\psi) \langle T^2_{+2}\rangle_3''],$

$\Upsilon(3) = -2\sqrt{2} \sin(2\theta) \cos(\theta) [\cos(\psi) \langle T^1\zeta\rangle_3 - \sin(\psi) \langle T^1\xi\rangle_3]$

$\qquad \times [\cos(\psi) \langle T^2_{+1}\rangle_3'' + \sin(\psi) \langle T^2_{+2}\rangle_3''].$

Reflection $(0, k, 0)$ with odd $k$ and $\xi_3 = (\eta_3 \times \zeta_3)$ is in the plane of scattering at the start of an azimuthal-angle scan. Dipoles $\langle T^1\xi\rangle_3 \equiv \langle T^1\eta\rangle_1$ and $\langle T^1\zeta\rangle_3 \equiv \langle T^1\xi\rangle_2$ on account of the triad axis of rotation symmetry along the c axis, and there is an additional quadrupole in $(\pi'\sigma)_3$ compared to $(\pi'\sigma)_1$ and $(\pi'\sigma)_2$. In consequence, $\Upsilon(3)$ is a function of $\sin(2\psi)$ and $\cos(2\psi)$, whereas $\Upsilon(1)$ and $\Upsilon(2)$ are functions of $\sin(2\psi)$ alone.

*Template* (4) All multipoles are purely real for template (4) and parity-even absorption events, for which $\sigma_\theta = (-1)^K$. Dipoles $\langle T^1\eta\rangle_1$ and $\langle T^1\xi\rangle_4$ are antiparallel and off-set from the dyad $\mathbf{a}_h$ by 30°. Notably, charge-like (even K) and magnetic (odd K) Bragg reflections overlap, and amplitudes for unrotated channels of polarization, $(\sigma'\sigma)_4$ and $(\pi'\pi)_4$, admit a scalar. It is proportional to the number of holes $N_o$ in the valence shell that accepts the photo-ejected photon, and very strong compared to lattice forbidden amplitudes created by small fractions of an electron [23, 24]. The scalar is absent in rotated channels of polarization, however, and for an E1-E1 event [23],

$(\pi'\sigma)_4 = (i/\sqrt{2}) [\cos(\theta) \sin(\psi) \langle T^1\xi\rangle_4 + \sin(\theta) \langle T^1\zeta\rangle_4]$

$\qquad - \cos(\theta) \cos(\psi) \langle T^2_{+1}\rangle'_4 - \sin(\theta) \sin(2\psi) \langle T^2_{+2}\rangle'_4.$ $\qquad (12)$

Reflection $(0, 0, l)$ with the reflection vector parallel to the c axis, as in all other cases already mentioned. The contribution to $(\pi'\sigma)_4$ from the dipole parallel to the c axis $\langle T^1\zeta\rangle_4$ is independent of the azimuthal angle, as expected. The start of an azimuthal angle scan has $(1, 0, 0)_h$ normal to the plane of scattering, as for template (1). Dipoles $\langle T^1\xi\rangle_4$ and $\langle T^1\zeta\rangle_4$ are contained in a plane spanned by $(1, 2, 0)_h$ and $(0, 0, 1)_h$ normal to a dyad axis of rotation symmetry. As with all our templates, charge-like and magnetic contributions to $(\pi'\sigma)_4$ differ in phase by 90° and an intensity $|(\pi'\sigma)_4|^2$ is a sum of squares.

The full expression for $\Upsilon(4)$ is long, simply because there are contributions from four amplitudes. Given that the chiral signature is likely dominated by the scalar charge we record the contribution to $\Upsilon(4)$ that is proportional to $N_o$, a relatively simple expression, namely,

$$\Upsilon(4) \approx N_o\, [\sin^3(\theta)\, \langle T^1_\zeta \rangle_4 - \cos^3(\theta) \sin(\psi)\, \langle T^1_\xi \rangle_4]. \tag{13}$$

At the Mn K-edge Miller indices are $l = 1, 3, 5, 7$. For $l = 7$ one finds $\sin^3(\theta) \approx 0.970$ and $\cos^3(\theta) \approx 0.003$, indicating that $\Upsilon(4)$ in Eq. (13) is independent of the azimuthal angle, to a good approximation.

Lastly, we give an example of diffraction enhanced an E2-E2 absorption event at the K edge (1s → 3d). Not surprisingly, contributions from dipoles and quadrupoles to E2-E2 and E1-E1 amplitudes are similar with respect to the azimuthal angle. We therefore include the distinguishing octupole $\langle T^3_Q \rangle_4$,

$$(\pi'\sigma)_4 \approx \cos(3\theta)\, [-i\sin(\psi)\, \langle T^1_\xi \rangle_4 + \sqrt{(30/7)} \cos(\psi)\, \langle T^2_{+1} \rangle'_4]$$

$$+ \sin(3\theta)\, [-i\langle T^1_\zeta \rangle_4 + \sqrt{(30/7)} \sin(2\psi)\, \langle T^2_{+2} \rangle'_4] \tag{14}$$

$$+ i\sin(\theta)\, [(3\cos^2(\theta) - 2)\, \langle T^3_0 \rangle_4 - \sqrt{30} \cos^2(\theta) \cos(2\psi)\, \langle T^3_{+2} \rangle'_4]$$

$$+ i\cos(\theta) \sin(\psi)\, [\sqrt{3}\, (\sin^2(\theta) + 1)\, \langle T^3_{+1} \rangle'_4 + \sqrt{5}\, (3\cos^2(\theta) - 2)\, \{1 + 2\cos(2\psi)\}\, \langle T^3_{+3} \rangle'_4].$$

The reflection is $(0, 0, l)$, as in the E1-E1 amplitude Eq. (12) to which Eq. (14) should be compared.

### VI. Magnetic neutron Bragg diffraction

The axial (parity-even) multipoles for neutron diffraction are denoted $\langle t^K_Q \rangle$ in Table II, and they are magnetic (time-odd, $\sigma_\theta = -1$). A scattering amplitude $\langle \mathbf{Q}_\perp \rangle$ generates an intensity $|\langle \mathbf{Q}_\perp \rangle|^2$ for unpolarized neutrons. In more detail, $\langle \mathbf{Q}_\perp \rangle = [\mathbf{e} \times (\langle \mathbf{Q} \rangle \times \mathbf{e})] = [\langle \mathbf{Q} \rangle - \mathbf{e}\,(\mathbf{e} \cdot \langle \mathbf{Q} \rangle)]$ with a unit vector $\mathbf{e} = \boldsymbol{\kappa}/\kappa$. The intermediate amplitude $\langle \mathbf{Q} \rangle$ is proportional to the magnetic moment $\langle \boldsymbol{\mu} \rangle$ in the forward direction of scattering, with $\langle \mathbf{Q} \rangle = \langle \boldsymbol{\mu} \rangle/2$ for $\kappa = 0$, apart from a numerical factor related to site multiplicity.

Magnetic neutron scattering amplitudes that follow are chosen to be purely real, and in phase with the nuclear scattering amplitude. Our templates designed for altermagnetism permit this property, and it is not universal; cf. Appendix. We continue to discuss basis forbidden reflections for templates (1), (2) and (3) for which the reflection vector parallel to the hexagonal c axis. Magnetic and nuclear amplitudes overlap in (4). For this symmetry we consider reflection vectors parallel to the c axis and $(1, -1, 0)_h$. The total intensity is $[|\langle \mathbf{Q}_\perp \rangle_a|^2 + |\langle \mathbf{Q}_\perp \rangle_c|^2]$, where subscripts a and c denote independent Mn ions in sites 2a and 2c in C2'/m'. Subtraction

of diffraction patterns gathered above and below the magnetic ordering temperature provided an estimate of the magnetic content [4].

Polarization analysis offers greater sensitivity to magnetic contributions, however. A polarization dependence of the neutron scattering can be described as a departure from unity of the ratio of the reflected intensities for incoming neutron beams of opposite polarization, i.e., $\mathbb{R} = (\mathbb{N} + \mathbb{Z})^2 /(\mathbb{N} - \mathbb{Z})^2$, where $\mathbb{N}$ and $\mathbb{Z}$ are total nuclear and magnetic amplitudes, respectively. For $\mathbb{R} \neq 1$ it is clear that neither $\mathbb{N}$ nor $\mathbb{Z}$ can be zero. Moreover, $\mathbb{N}$ and $\mathbb{Z}$ must have like phases [28, 29]. More generally, a fraction $\propto \{(1/2) (1 + P^2) |\langle \mathbf{Q}_\perp \rangle|^2 - |\mathbf{P} \cdot \langle \mathbf{Q}_\perp \rangle|^2\}$ of neutrons participate in events that change (flip) the neutron spin orientation, where $\mathbf{P}$ is the primary polarization. The assumption of perfect polarization $(\mathbf{P} \cdot \mathbf{P}) = 1$ yields a spin-flip signal [31, 32],

$$\text{SF} = [|\langle \mathbf{Q}_\perp \rangle|^2 - |\mathbf{P} \cdot \langle \mathbf{Q}_\perp \rangle|^2]. \tag{15}$$

Evidently, all scattering is spin-flip when $\mathbf{P}$ and $\mathbf{e}$ are aligned since $\mathbf{e} \cdot \langle \mathbf{Q}_\perp \rangle = 0$.

Magnetic multipoles in neutron diffraction depend on the magnitude of the reflection vector, $\kappa$. The dipole $\langle \mathbf{t}^1 \rangle$ contains standard radial integrals $\langle j_0(\kappa) \rangle$ and $\langle j_2(\kappa) \rangle$ shown in Fig. 4, with $\langle j_0(0) \rangle = 1$ and $\langle j_2(0) \rangle = 0$ [32]. An approximation to $\langle \mathbf{t}^1 \rangle$ suitable for a transition-metal ion is [25, 26],

$$\langle \mathbf{t}^1 \rangle \approx (\langle \boldsymbol{\mu} \rangle/3) [\langle j_0(\kappa) \rangle + \langle j_2(\kappa) \rangle (g - 2)/g]. \tag{16}$$

Here, the magnetic moment $\langle \boldsymbol{\mu} \rangle = g \langle \mathbf{S} \rangle$ and the orbital moment $\langle \mathbf{L} \rangle = (g - 2) \langle \mathbf{S} \rangle$. The coefficient of $\langle \mathbf{L} \rangle$ is approximate, while $\langle \mathbf{t}^1 \rangle = (1/3) \langle 2\mathbf{S} + \mathbf{L} \rangle$ for $\kappa \to 0$ is an exact result. Higher order multipoles with even rank depend on the electronic position operator $\mathbf{n}$. The equivalent operator $[(\mathbf{S} \times \mathbf{n}) \mathbf{n}]$ for $\mathbf{t}^2$ shows that actually the quadrupole measures the correlation between the spin anapole $(\mathbf{S} \times \mathbf{n})$ and orbital degrees of freedom [26]. The quadrupole $\langle \mathbf{t}^2 \rangle$ is proportional to $\langle j_2(\kappa) \rangle$ in Fig. 4.

Magnetic neutron multipoles with an even rank do not exist for magnetic states derived from a J-state, instead, a ground state must possess two or more J-states for $\langle \mathbf{t}^2 \rangle$ non-zero [26]. A single J-state is likely at odds with the basic premise of altermagnetism because it is an outcome of a strong spin-orbit coupling [10-15]. Indeed, a Landau theory in the extreme zero spin-orbit coupling limit has been shown to capture the essence of altermagnetism [30]. Quadrupoles and hexadecapoles (K = 4) are permitted by all our templates.

The following expressions for magnetic neutron scattering amplitudes are correct for dipoles and quadrupoles, and we set multipoles of rank K = 3, 4, 5 aside [26]. One finds $\langle \mathbf{Q}_\perp \rangle = \langle \mathbf{Q} \rangle$ in symmetries (1), (2) and (3) because $\mathbf{e} \cdot \langle \mathbf{Q} \rangle = 0$.

*Template* (1) Reflections (0, 0, $l$) with odd $l$ and $\langle \mathbf{Q}_\perp \rangle_1 = (0, \langle Q_\eta \rangle_1, 0)$ with,

$$\langle Q_\eta \rangle_1 \approx 2 [3 \langle t^1_\eta \rangle_1 - 2\sqrt{3} \langle t^2_{+1} \rangle'_1]. \tag{17}$$

Referring to Table I, the dipole is aligned with $(1, 2, 0)_h$.

*Template* (2) Reflections $(0, 0, l)$ with odd $l$ and $\langle \mathbf{Q}_\perp \rangle_2 = (\langle Q_\xi \rangle_2, 0, 0)$ with,

$$\langle Q_\xi \rangle_2 \approx 2\, [3\, \langle t^1_\xi \rangle_2 + 2\sqrt{3}\, \langle t^2_{+1} \rangle''_2]. \tag{18}$$

*Template* (3) Reflections $(0, k, 0)$ with odd $k$ and,

$$\langle Q_\xi \rangle_3 \approx 3\, \langle t^1_\xi \rangle_3 - 2\sqrt{3}\, \langle t^2_{+1} \rangle''_3,\ \langle Q_\eta \rangle_3 = 0,\ \langle Q_\zeta \rangle_3 \approx 3\, \langle t^1_\zeta \rangle_3 - 2\sqrt{3}\, \langle t^2_{+2} \rangle''_3. \tag{19}$$

*Template* (4) Reflections $(0, 0, l)$ and all $l$. For Mn ions in sites 2a and 2c the neutron diffraction amplitude $\langle \mathbf{Q} \rangle_4 \approx (\langle Q_\xi \rangle_4, 0, 3\langle t^1_\zeta \rangle_4)$ and $\langle \mathbf{Q}_\perp \rangle_4 \approx (\langle Q_\xi \rangle_4, 0, 0)$ with,

$$\langle Q_\xi \rangle_4 \approx 3\, \langle t^1_\xi \rangle_4 + 2\sqrt{3}\, \langle t^2_{+1} \rangle''_4. \tag{20}$$

Reflections $(0, k, 0)$ and even $k$, to respect C-centring, with $\mathbf{e} = (1/2)\, (\boldsymbol{\xi} + \sqrt{3}\, \boldsymbol{\eta}) \propto (1, -1, 0)_h$. One finds,

$$\langle Q_{\perp\xi} \rangle_4 \approx (3/2)\, [(3/2)\, \langle t^1_\xi \rangle_4 - \sqrt{3}\, \langle t^2_{+1} \rangle''_4],\ \langle Q_{\perp\eta} \rangle_4 \approx (3/2)\, [-(1/2)\, \sqrt{3}\, \langle t^1_\xi \rangle_4 + \langle t^2_{+1} \rangle''_4],$$

$$\langle Q_{\perp\zeta} \rangle_4 \approx 3\, \langle t^1_\zeta \rangle_4 - \sqrt{3}\, \langle t^2_{+2} \rangle''_4. \tag{21}$$

Recall that local coordinates for dipoles in template (4) are $\boldsymbol{\xi}_4 = -(1, 2, 0)_h$ and $\boldsymbol{\zeta}_4 = (0, 0, 1)_h$, with $(1, 2, 0)_h \equiv (1, -1, 0)_h$.

## VII. Conclusions

In summary, templates specified in Table I possess magnetic symmetries suitable for hexagonal manganese telluride ($\alpha$-MnTe, Fig. 1), and differ with respect to the orientation of Mn dipoles in the basal plane and canting out of the plane. Moreover, they satisfy an accepted definition of an altermagnet [12]. Our motivation is the prospect of matching a measured Bragg diffraction pattern with a template. Inherent symmetry requires neutron diffraction amplitudes to be purely real (or purely imaginary), and polarization analysis is available to separate magnetic signals and nuclear signals when the contributions overlap [29]. Identical symmetry requires magnetic and charge-like (T & T [20]) contributions to x-ray amplitudes to be 90º out of phase. Meaning that circular polarization in the primary x-ray beam is rotated upon diffraction [24]. The mentioned attributes of our diffraction amplitudes are not universal, cf. Appendix. But they are attributes expected for diffraction by an altermagnet. Likewise, neutron multipoles with an even rank are not supressed by a spin-orbit coupling.

Other properties that can be tested in resonant x-ray diffraction include zero diffraction in the unrotated $\sigma$ polarization channel (Fig. 3) for templates (1), (2) and (3) in Table I. Templates (1) and (4) permit ferromagnetism as a secondary order parameter and magnetic circular dichroism (XMCD); dipoles in (4) are depicted in Fig. 2. They also allow dipoles in the basal plane off-set by 30º from the dyad axis of rotation symmetry. By contrast, the dipole in template (2) is confined to the dyad. Templates (1), (2) and (3) possess reflections that are lattice forbidden, i.e., they do not exist in nuclear or Thomson scattering by the parent structure.

Our templates are centrosymmetric, and possess long-range antiferromagnetic order with a propagation vector $\mathbf{k} = (0, 0, 0)$. Anti-inversion, requiring invariance with respect to the product of parity and the reversal of time $\overline{1}'$, is absent in all magnetic crystal classes listed in

Table I. Landau free-energies include a contribution HEE, where H and E denote magnetic and electric fields, and a piezomagnetic effect is permitted [6]. Nineteen of the 122 magnetic crystal classes include anti-inversion, together with a Landau free-energy EH and a linear magnetoelectric effect. Chromium sesquioxide $Cr_2O_3$ is an epitome material (magnetic crystal class $\bar{3}'$m'). Today, parity-time (PT)-symmetric antiferromagnets attract attention because of their nonlinear responses, e.g., a second-order Hall effect [34]. CuMnAs depicted in Fig. 5 is cited in this context, and it is the subject of an Appendix by way of a contrast to properties of our templates [35, 36]. Specifically, magnetic and charge-like contributions to x-ray amplitudes for CuMnAs possess the same phase and a chiral signature is forbidden. The orthorhombic semimetal supports antiferromagnetic **k** = (0, 0, 0) long-range order described by magnetic symmetry Pn'ma (No. 62.443 [16]), which belongs to the magnetic crystal class m'mm. Properties of this crystal class differ from the four listed in Table I for our templates with respect to just two bulk effects; a linear magnetoelectric effect is allowed in m'mm, and the piezomagnetic effect is forbidden.

**Acknowledgements**. Dr Urs Staub advised on the feasibility of the proposed resonant x-ray Bragg diffraction using Mn edges. Dr Sarnjeet Dhesi discussed experimental findings mentioned in Ref. [9].

## Appendix

Emmanouilidou *et al*. [36] found the magnetism of CuMnAs very sensitive to the stoichiometry of the Cu and Mn sites. They make the statement "While $Cu_{0.95}$MnAs is a commensurate antiferromagnet below 360 K with a propagation vector of **k** = (0, 0, 0), $Cu_{0.98}Mn_{0.96}$As undergoes a second-order paramagnetic to incommensurate antiferromagnetic phase transition at 320 K with **k** = (0.1, 0, 0), followed by a second-order incommensurate to commensurate antiferromagnetic phase transition at 230K". Here we explore a commensurate antiferromagnet CuMnAs ($Mn^{2+}$, $3d^5$) depicted in Fig. 5.

Manganese ions occupy sites 4c in Pn'ma (magnetic crystal class m'mm) that are not centres of inversion symmetry, unlike sites used by Mn ions in our templates. Site symmetry demands $[\sigma_\pi (-1)^{K+Q} \langle O^K_{-Q}\rangle] = \langle O^K_Q\rangle$, with $(-1)^Q \langle O^K_{-Q}\rangle = \langle O^K_Q\rangle^*$. The PT-symmetric Pn'ma electronic structure factor,

$$\Psi^K_Q(4c) = \langle O^K_Q\rangle \{\alpha\gamma + (\alpha\gamma)^* \sigma_\theta \sigma_\pi (-1)^k \quad (A1)$$

$$+ \sigma_\pi (-1)^{h+l+Q} [\alpha\gamma^* + (\alpha\gamma^*)^* \sigma_\theta \sigma_\pi (-1)^k]\}.$$

Spatial phase factors $\alpha = \exp(i2\pi hx)$ and $\gamma = \exp(i2\pi lz)$, where x ≈ 0.460 and z ≈ 0.677 are general coordinates. As with MnTe, cell lengths a ≈ 6.57 Å, b ≈ 3.86 Å, c ≈ 7.30 Å of CuMnAs are too small for resonant x-ray diffraction at Mn $L_{2,3}$ absorption edges [36]. Extinction rules for lattice reflections are derived from $\Psi^K_0(4c)$ evaluated with even K and $\sigma_\theta = \sigma_\pi = +1$, namely, (h, k, 0) with even h, and (0, k, l) with even k + l.

For lattice forbidden reflections ($h$, 0, 0) with odd $h$, and an E1-E1 absorption event (1s → 4p) the x-ray amplitudes are ($\sigma'\sigma$) = 0,

$$(\pi'\pi) = 2\sqrt{2} \sin(2\theta) \sin(\psi) \, \alpha'' \, \langle T^1_b \rangle,$$

$$(\pi'\sigma) = 2 \cos(\theta) \cos(\psi) [\sqrt{2}\alpha'' \langle T^1_b \rangle - 2\alpha' \langle T^2_{+1} \rangle']. \quad (A2)$$

X-ray amplitudes are purely real or zero and the chiral signature is zero, as anticipated in Section VII. The configuration of axial dipoles $\langle T^1_b \rangle$ is illustrated in Fig. 5. At the start of an azimuthal angle scan $\psi = 0$ the crystal b axis is in the plane of scattering, Fig. 3.

Dirac multipoles $\langle G^K_Q \rangle$ in Table II diffract x-rays in the $\sigma'\sigma$ channel of polarization for reflections ($h$, 0, 0) with odd $h$, unlike the foregoing result for parity-even E1-E1. For an E1-E2 (1s → 3d, 4p) event [23],

$$(\sigma'\sigma) = (4/5) \, \alpha' \sqrt{3} \cos(\theta) \sin(\psi) \{ - \langle G^1_c \rangle + (1/3) \sqrt{10} \, \langle G^2_{+2} \rangle''$$

$$+ \sqrt{(2/3)} [1 - 5 \cos^2(\psi)] \langle G^3_0 \rangle + (2/3) \sqrt{5} [1 - 3 \cos^2(\psi)] \langle G^3_{+2} \rangle' \}. \quad (A3)$$

The remaining three E1-E2 x-ray amplitudes are also purely real. The Dirac dipole (anapole) parallel to the crystal a axis $\langle G^1_a \rangle$ does not contribute to ($\sigma'\sigma$) at ($h$, 0, 0) with odd $h$.

All E1-E1 amplitudes are zero for (0, $k$, 0) with odd $k$. Likewise for diffraction enhanced by the parity-odd E1-E2 event, and neutron diffraction that we consider next.

Neutron diffraction by CuMnAs is created by axial $\langle t^K_Q \rangle$ and Dirac $\langle g^K_Q \rangle$ multipoles that feature in Table II. Scattering amplitudes are labelled by superscripts $^{(\pm)}$ using + (−) for axial (Dirac) amplitudes. Continuing a study of lattice forbidden reflections ($h$, 0, 0) with odd $h$, we find $\langle \mathbf{Q}_\perp \rangle^{(+)} = (0, \langle Q_b \rangle^{(+)}, 0)$ with,

$$\langle Q_b \rangle^{(+)} \approx 4i\alpha'' [(3/2) \langle t^1_b \rangle + \sqrt{3} \langle t^2_{+1} \rangle']. \quad (A4)$$

Eq. (16) provides an estimate of the dipole parallel to the b axis in Eq. (A4), using radial integrals in Fig. 4. The anapole in neutron diffraction $\langle \mathbf{g}^1 \rangle$ is a sum of three multipoles, including a spin anapole ($\mathbf{S} \times \mathbf{n}$) and an orbital analogue $[(\mathbf{L} \times \mathbf{n}) - (\mathbf{n} \times \mathbf{L})]$ [18, 26]. Associated radial integrals for $Mn^{2+}$ are displayed in Fig. 3 of Ref. [38]. The Dirac scattering amplitude for reflections ($h$, 0, 0) with odd $h$ has a component parallel to the crystal b axis, namely,

$$\langle Q_b \rangle^{(-)} \approx 4i\alpha' [- \langle g^1_c \rangle + (3/\sqrt{5}) \langle g^2_{+2} \rangle''], \quad (A5)$$

and $\langle \mathbf{Q}_\perp \rangle^{(-)} = (0, \langle Q_b \rangle^{(-)}, 0)$. Notably, $\langle \mathbf{Q}_\perp \rangle^{(\pm)}$ are purely imaginary. Dirac and axial amplitudes are correct at the level of quadrupoles, as in the main text.

Axial and polar (Dirac) neutron diffraction amplitudes are zero for (0, 0, $l$) with odd $l$. Two of the four x-ray amplitudes are zero for both E1-E1 and E1-E2 enhancements, namely, amplitudes diagonal in polarization states (σ'σ) = (π'π) = 0. An azimuthal angle scan starts with the b axis in the plane of scattering and directed along − y in Fig. 3. For the parity-even event E1-E1,

$$(\pi'\sigma) = (\sigma'\pi) = -4\,\gamma'\cos(\theta)\cos(\psi)\,\langle T^2_{+1}\rangle'. \quad (0, 0, l) \text{ odd } l. \qquad (A6)$$

In other words, E1-E1 diffraction at these reflections is exclusively T & T scattering [20]. For enhancement by an E1-E2 event [23],

$$(\pi'\sigma) = -(\sigma'\pi) = 2\,\gamma'\,\sqrt{(2/15)}\cos^2(\theta)\sin(2\psi)$$

$$\times [-\langle G^2_{+2}\rangle'' + 2\sqrt{2}\,\langle G^3_{+2}\rangle']. \quad (0, 0, l) \text{ odd } l. \qquad (A7)$$

No multipoles are neglected in (A6) and (A7).

**References**.

[38] S. W. Lovesey, Phys. Rev. B **107**, 224410 (2023).

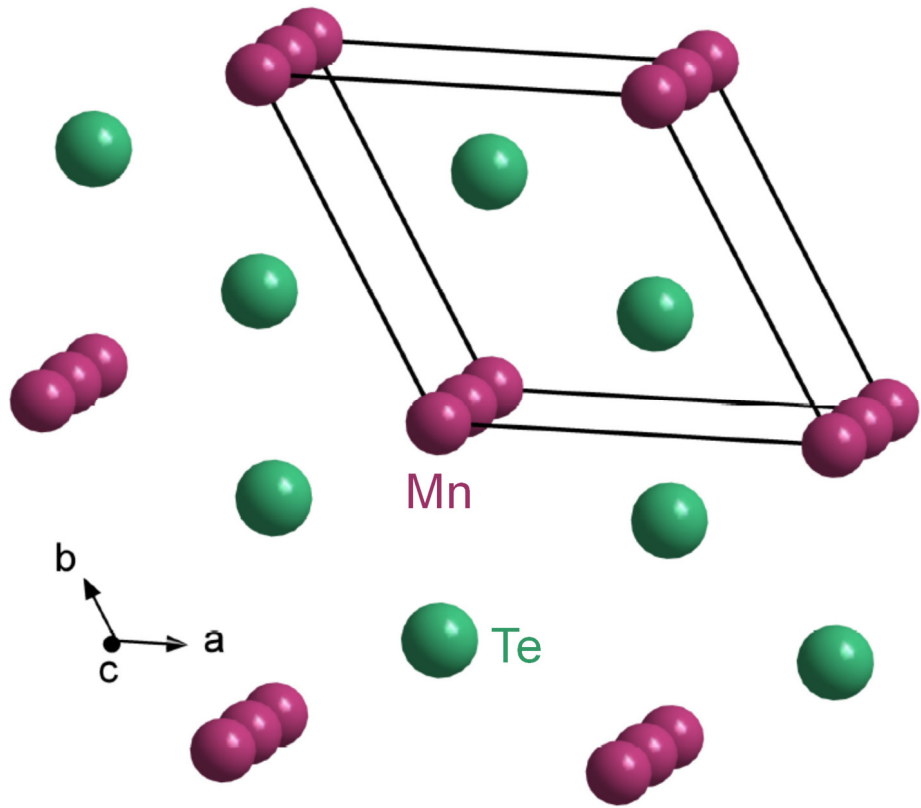

FIG. 1. Chemical structure of MnTe.

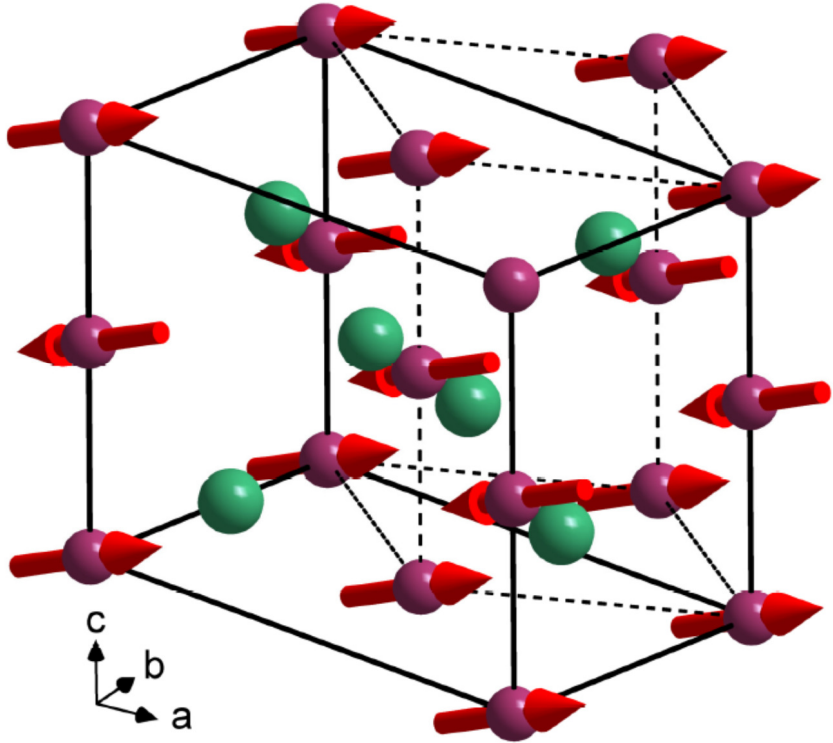

FIG. 2. Configuration of Mn dipoles in the template with symmetry C2'/m' (No. 12. 62) labelled (4) in the text and Table I.

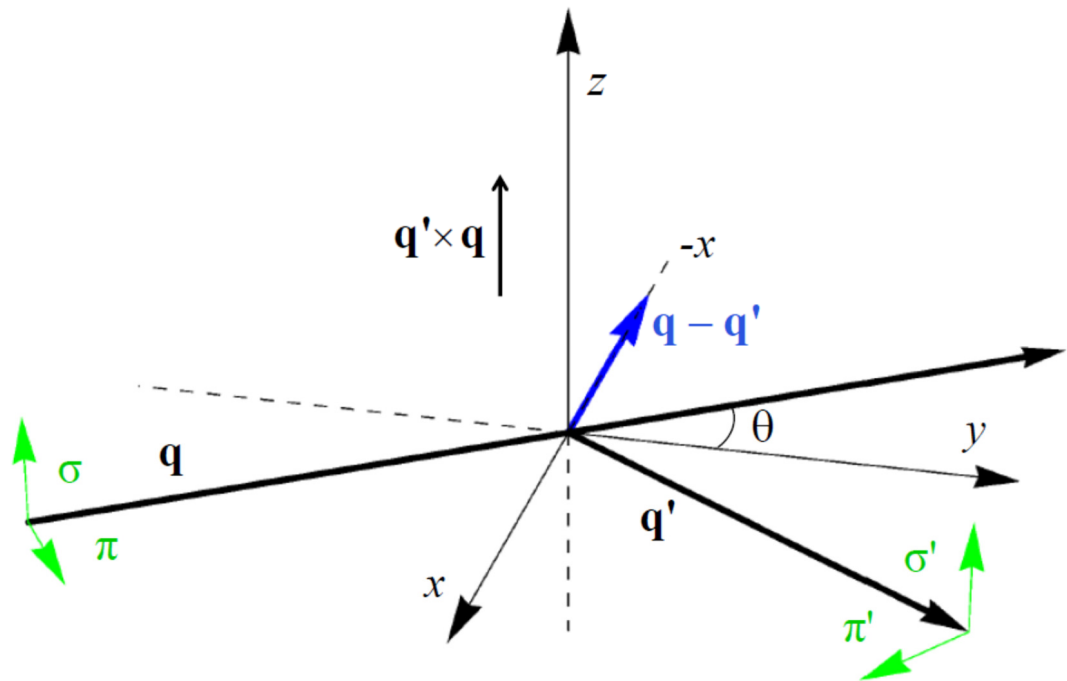

FIG. 3. Primary ($\sigma$, $\pi$) and secondary ($\sigma'$, $\pi'$) states of polarization. Corresponding wavevectors **q** and **q'** subtend an angle $2\theta$. The Bragg condition for diffraction is met when **q** − **q'** coincides with a reflection vector ($h$, $k$, $l$) of the reciprocal lattice. Crystal vectors that define local axes ($\xi$, $\eta$, $\zeta$) and the depicted Cartesian (x, y, z) coincide in the nominal setting of the crystal.

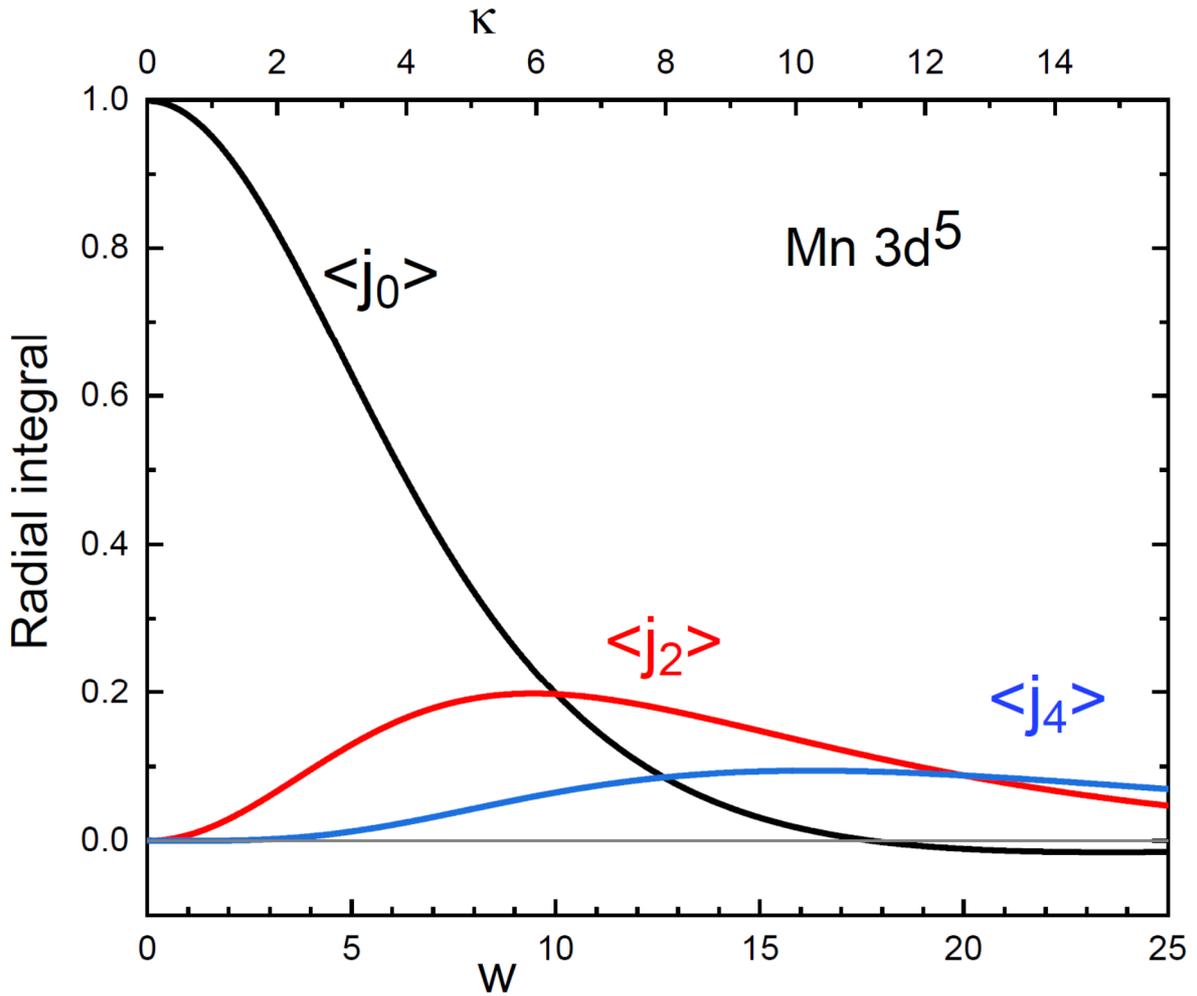

FIG. 4. Radial integrals $\langle j_0 \rangle$ (black), $\langle j_2 \rangle$ (red), and $\langle j_4 \rangle$ (blue) for $Mn^{2+}$ ($3d^5$) calculated using Cowan's code [33]. The dimensionless parameter w and the magnitude of the reflection vector $\kappa$ are related by the Bohr radius, namely, $\kappa = w/3a_o$. Also, $\kappa = 4\pi s$ with $s = \sin(\theta)/\lambda$ and $\lambda$ the neutron wavelength (cf. Section 6.3.1, [27]).

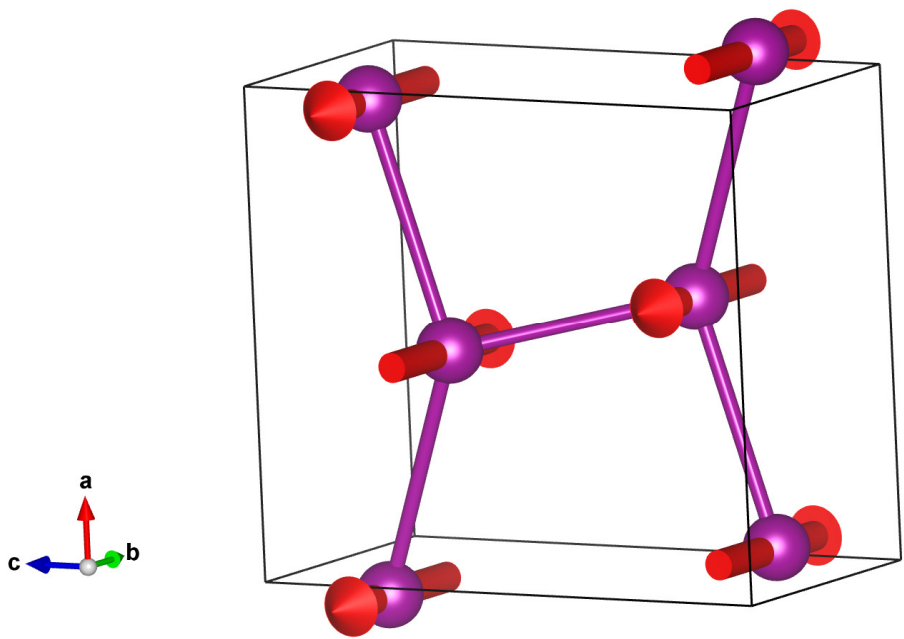

FIG. 5. Configuration of Mn axial dipoles in CuMnAs (magnetic crystal class m'mm) with magnetic symmetry Pn'ma (No. 62.443) [36]. Reproduced from MAGNDATA [37].

**TABLE I**. Templates for magnetic symmetry in hexagonal MnTe below a Néel temperature 307 - 310 K. The primary order parameter is an antiferromagnetic alignment between dipoles of the two Mn ions. Magnetic groups and Mn site symmetries are centrosymmetric ($\mathfrak{I}$ denotes spatial inversion). The magnetic order does not break translation symmetry, and anti-inversion ($\bar{1}'$) is absent in magnetic crystal classes. All crystal classes permit a piezomagnetic effect and forbid a linear magnetoelectric effect. Ferromagnetism as a secondary order parameter is allowed (Y) in (1), (3) and (4) and forbidden (N) in (2). The hexagonal lattice vector $(1, 0, 0)_h$ is parallel to a dyad axis of rotation symmetry, and $(1, 0, 0)_h$ and $(1, -1, 0)_h$ subtend an angle of 30°. The volume of a unit cell $v_o = (1/2)\, a^2 c \sqrt{3}$ for (3) and (4), and the volume of an orthorhombic unit cell = $2v_o$.

| **Template** | (1) | (2) | (3) | (4) |
|---|---|---|---|---|
| *Magnetic group* | Cm'c'm | Cmcm | P2$_1$/m | C2'/m' |
| *Crystal class* | m'm'm | mmm | 2/m | 2'/m' |
| *Mn site symmetry* | 2'/m' | 2/m | $\mathfrak{I}$ | 2'/m' |
| *Ferromagnetism* | Y | N | Y | Y |
| *Dipole direction in the* (ab) *plane* | $(1, -1, 0)_h$ | $(1, 0, 0)_h$ | general | $(1, -1, 0)_h$ |

**TABLE II.** A generic multipole $\langle O^K_Q \rangle$ has integer rank K and $(2K + 1)$ projections Q in the interval $-K \leq Q \leq K$. Angular brackets $\langle ... \rangle$ denote the expectation value, or time average, of the enclosed spherical tensor operator. Parity ($\sigma_\pi$) and time ($\sigma_\theta$) signatures = $\pm 1$, e.g., $\langle t^K_Q \rangle$ for magnetic neutron diffraction is parity-even ($\sigma_\pi = +1$) and time-odd ($\sigma_\theta = -1$). Manganese ions in MnTe occupy sites that are centres of inversion symmetry resulting in $\sigma_\pi = +1$ for all multipoles. Not so for CuMnAs, leading to a need for Mn Dirac multipoles $\langle g^K_Q \rangle$ and $\langle G^K_Q \rangle$ with $\sigma_\theta \sigma_\pi = +1$.

| **Signature** | $\sigma_\pi$ | $\sigma_\theta$ |
|---|---|---|
| *Neutrons* | | |
| $\langle t^K_Q \rangle$ | +1 | −1 |
| $\langle g^K_Q \rangle$ | −1 | −1 |
| | | |
| *Photons* | | |
| $\langle T^K_Q \rangle$ | +1 | $(-1)^K$ |
| $\langle G^K_Q \rangle$ | −1 | −1 |